# Optimal Design of Thin-film Plasmonic Solar Cells using Differential Evolution Optimization Algorithms

Ankit Vora
Dept. of Elec. &
Comp. Eng.
Michigan Tech. Univ.
Houghton, MI, USA
avora@mtu.edu

Satyadhar Joshi
Grad. School of Tech.
Touro College
New York, NY, USA

Arun Matai
Dept. of Elec. &
Comm. Eng.
Acropolis Tech.
Campus
Indore, MP, INDIA

Joshua M. Pearce
Dept. of Elec. &
Nanoeng.
School of Elec. Eng.
Aalto Univ.
Espoo, Finland
Dept. of Elec. &
Comp. Eng.
Dept. of Mat. Sci. &
Eng.
Michigan Tech. Univ.
Houghton, MI, USA

Durdu Guney
Dept. of Elec. &
Comp. Eng.
Michigan Tech. Univ.
Houghton, MI, USA

*Abstract*— An approach using a differential evolution (DE) optimization algorithm is proposed to optimize design parameters for improving the optical absorption efficiency of plasmonic solar cells (PSC). This approach is based on formulating the parameters extraction as a search and optimization process in order to maximize the optical absorption in the PSC. Determining the physical parameters of three-dimensional (3-D) PSC is critical for designing and estimating their performance, however, due to the complex design of the PSC, parameters extraction is time and calculation intensive. In this paper, this technique is demonstrated for the case of commercial thin-film hydrogenated amorphous silicon (a-Si:H) solar photovoltaic cells enhanced through patterned silver nano-disk plasmonic structures. The DE optimization of PSC structures was performed to execute a real-time parameter search and optimization. The predicted optical enhancement (OE) in optical absorption in the active layer of the PSC for AM-1.5 solar spectrum was found to be over 19.45% higher compared to the reference cells. The proposed technique offers higher accuracy and automates the tuning of control parameters of PSC in a time-efficient manner.

## I. INTRODUCTION

Solar energy is envisioned to be an essential source of energy in the future [1]. In particular, the solar photovoltaic (PV) cells, which converts solar energy directly into electrical energy, are becoming a viable renewable energy source due to its continually declining levelized cost of electricity [2]. Hydrogenated amorphous silicon (a-Si:H) is readily available, earth abundant and inexpensive solar cell material which has the potential to assist in generating clean, sustainable energy at the terawatt scale [1]. However, the ultimate technological challenge confronting the a-Si:H PV cells is light-induced degradation of electrical performance known as the Staebler-Wronski effect (SWE) [3,4]. When a-Si:H based PV cells are exposed to light, there is a decrease in their efficiency until a saturation value is reached (degraded steady state or DSS) due to an increase in the density of multiple types of defect states [4]. However, recent advances in the field of plasmonics provide a new method to enhance the optical absorption in a-Si:H PV devices and combat the detrimental effects of SWE [5-16]. The conventional thin-film PSC consist of patterned arrays of nanostructured plasmonic resonators on the top or the bottom of the solar PV cell [5-16]. Typically, the surface plasmon resonances manifested by plasmonic nanostructures demonstrate a strong relationship to the physical and material attributes of the surrounding medium (shape, size and the dielectric properties). In order to estimate the thin-film PSC performance, the extraction of optimized physical parameters of the device is essential. However, due to the increased number of parameters pertaining to the sensitive nanostructured plasmonic resonators, the optimization of physical parameters of the cell becomes challenging both temporally as well as computationally. An obvious question arises: how to optimize the physical parameters of the PSC in order to maximize the device power output. In this paper, we present a combination method of the fully vectorial finite element based optical modeling of solar PV cells to compute the optical absorption characteristics in the active layer of the solar cells along with the differential evolution algorithm for optimizing the physical parameters of the device.

*A. Overview of Differential Evolution Algorithms*

The DE optimization algorithm, formulated by Storn and Price in 1997 [17], is a simple and efficient black-box optimization approach which has three major advantages: i) global minimum extraction irrespective of the initial parameter values, ii) few control parameters required, and iii) rapid convergence. Black-box (or derivative-free) optimization finds the global minimum of a function when the analytical form is unknown, and therefore derivatives computations cannot be performed to minimize it. A bound constrained problem for black-box global optimization can be defined as follows:

$$\min f(X), \ X = [x_1,...,x_D]; \ S.t. \ x_j \in [a_j, b_j] \quad (1)$$
$$j = 1, 2, ..., D$$

where $f$ is the objective function to be minimized, $X$ is the decision vector comprising of $D$ parameters to be optimized, and $a_j$ and $b_j$ are the lower and upper bounded constraint for each parameter, respectively. A set of $D$ optimization parameters is defined as an individual. The DE is a population-based direct search optimization algorithm, which evolves a population of $NP$ individuals of dimension $D$ towards the global minimum by iteratively enhancing the differential among individuals. The population size is not altered during the evolution process, and the initial population is randomly chosen from a uniformly distributed parameters search space.

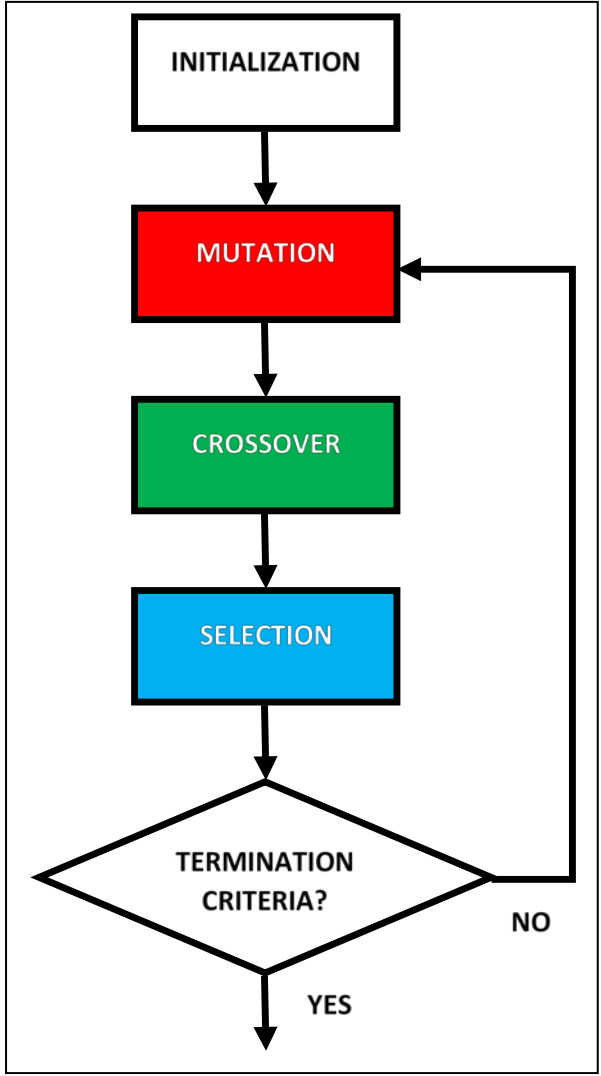

Fig. 1 A generalized block diagram of DE optimization algorithm.

The block diagram representing the operation of DE optimizing algorithm is depicted in Fig 1. The DE performs on two populations; old generation and a new generation of the previous population. After initialization, DE reiterates three operations on the population: mutation, crossover, and selection [17]. For each generation of population, the current population individuals become target vectors, and the mutation operator generates a mutant vector for each target vector. A new vector, termed as trial vector, is created by the crossover operator through a combination of parameters of the mutant and the target vectors [17]. The selection operator evaluates the trial vector for its fitness, and if this results in an improved fitness value compared to the target vector's fitness value, then the target vector is swapped with the trial vector by the selection operator in the successive generation. This process of mutation, crossover, and selection is reiterated until satisfying some specific termination criteria.

In mathematical notation, the population is represented by:

$$P_{X,G} = (X_{i,G}),\ i=1,...,NP,\ G=1,...,G_{max}$$
$$X_{i,G} = (x_{j,i,G}),\ j=1,...,D \quad (2)$$

where $P$ is the population and $X$ is the $i^{th}$ individual of the generation $G$, and the population size is $NP$. The dimension of the vector $X$ is $D$, which are the parameters to be optimized, and $x_{j,i,G}$ is the $j^{th}$ parameter of the $i^{th}$ individual of the generation $G$.

Initialization: For each parameter $j$ with $a_j$ and $b_j$ as the lower and upper bounds respectively, initial parameter values are chosen randomly with uniform distribution in the interval $[a_j,b_j]$:

$$x_{i,j} = a_j + rand_j \cdot (b_j - a_j) \quad (3)$$

where $rand_j$ is a random number being uniformly distributed between [0,1], which generates a new value for each decision parameter $x_j$ [17].

Mutation: For each target vector $X_{i,G}$ in the population, the mutation operation generates a mutant vector $V_{i,G}$ as follows [17]:

$$V_{i,G} = X_{r_1,G} + F \cdot (X_{r_2,G} - X_{r_3,G})$$
$$r_1, r_2, r_3, i \in \{1,...,NP\},\ r_1 \neq r_2 \neq r_3 \neq i \quad (4)$$

where the indices $r_1$, $r_2$, and $r_3$ are randomly chosen integers within the range [1,NP]. $F$ is a mutation scaling factor and a constant in the range from (0, 2) to control the augmentation of the difference vector $(X_{r2,G} - X_{r3,G})$, and any off-bound component of a mutant vector is regenerated again using equation (3). A larger scaling factor increases the search radius but may slow down the convergence of the algorithm.

Crossover: The target vector $X$ is crossbred with the mutated vector $V$ to yield the trial vector $U$ to complement the differential mutation search strategy further using the following scheme [17]:

$$U_{i,G} = \begin{cases} v_{j,i,G}, & if\ (rand_j(0,1) \leq C_r\ or\ j = j_{rand}) \\ x_{j,i,G}, & otherwise \end{cases} \quad (5)$$
$$j=1,...,D,\ and\ i=1,...,NP$$

where $rand_j$ is a random number being uniformly distributed between [0,1], $C_r$ is the user-defined crossover probability, a constant within the range [0,1], which regulates the portion of parameter values getting copied to the donor vector from the mutant vector, and $j_{rand}$ is a random integer within the range [1,D].

Selection: The target vector $X$ and the trial vector $U$ are compared after evaluation, and the selection operator admits the vector with better fitness value to the next generation. After the termination of DE optimization algorithm, the selection operation will generate the optimized parameters of the $f(X)$ function. The selection operation in DE can be expressed as follows [17]:

$$X_{i,G+1} = \begin{cases} U_{i,G}, & \text{if } f(U_{i,G}) \leq f(X_{i,G}) \\ X_{i,G}, & \text{otherwise} \end{cases} \quad (6)$$

$i \in [1, NP]$

In this work, we investigated the application of the DE optimization algorithm to optimize the a-Si:H based PSC. Due to the complex design of the cells, a black-box optimization algorithm is best suited to automate the design procedure. Our previous work [9] has already shown promising results. However, in that work, the cell optimization was performed manually by investigating all the independent parameters in a controlled manner. Here, we investigate the effectiveness of DE optimization to outperform the already reported promising design.

## II. METHODS

The a-Si:H thin-film PSC optimized by DE algorithm in this paper is based on our previous work of developing the patterned plasmonic nano-disks based a-Si:H solar cells [9]. The thin-film plasmonic a-Si:H solar cell used in this study is depicted in Fig. 2 along with the reference structure for comparison. The optical responses of the reference and PSC were obtained through COMSOL Multiphysics v5.3b evaluated in RF module in frequency domain coupled with MATLAB R2016b. The details of the simulations are described in our previously published literature [9]. The absorbance in individual layers of the solar cells was evaluated from the in-built power loss density function in COMSOL and the absorbance in the active layer (i-a-Si:H layer) of solar cells was used to evaluate the absorbed power density in the PSC and the reference structure [12]. The absorbed power density in the active layer of solar cells for incident reference solar spectrum (NREL AM 1.5) was calculated using the equation [12]:

$$P_{i-a-Si:H} = \int A(\lambda) E_{AM1.5}(\lambda) d\lambda \quad (7)$$

where $P_{i-a-Si:H}$ is the absorbed power density (measured in W/m$^2$) in the i-a-Si:H layer for AM 1.5 reference solar spectrum, $A(\lambda)$ is the absorbance in i-a-Si:H layer as a function of wavelength, and $E_{AM1.5(\lambda)}$ is the spectral irradiance as a function of wavelength obtained from NREL as described in [12]. The optical enhancement $(OE)$ is defined and evaluated using the expression:

$$OE = \left( \frac{P_{i-a-Si:H(PSC)}}{P_{i-a-Si:H(\text{Re}f)}} - 1 \right) \times 100 \quad (8)$$

where the subscript '*PSC*' and '*Ref*' denote the PSC and the reference solar cells, respectively.

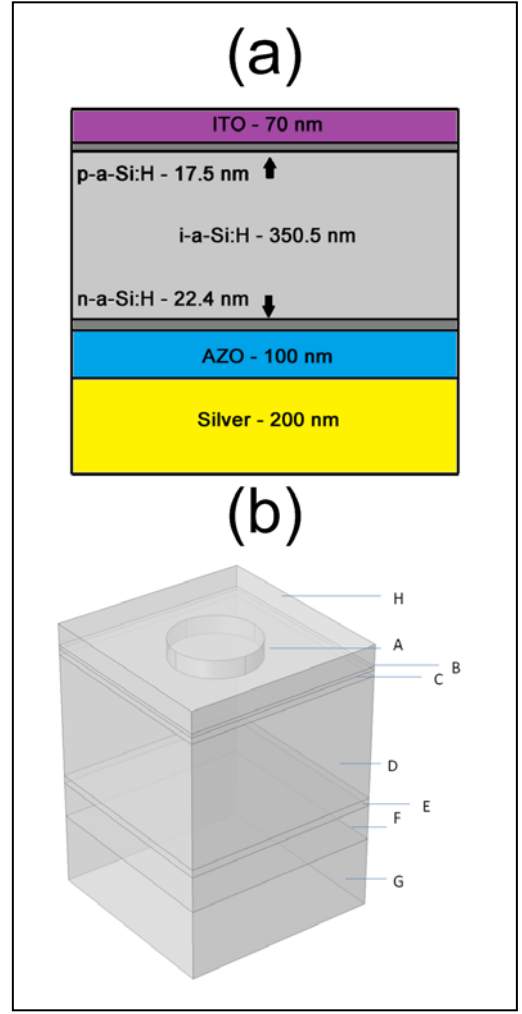

Fig. 2 (a) Reference structure, the top layer is ITO of 70 nm thickness, followed by p-a-Si:H of 17.5 nm thickness, then i-a-Si:H of 350.5nm thickness, after that n-a-Si:H of 22.5 nm thickness, followed by AZO of 100nm thickness and the last is a silver ground plate of 200 nm. (b) a-Si:H thin-film PSC A) nano-disk, diameter: 230 nm, height: 51 nm B) ITO layer, height: 10 nm, C) p-layer, height: 17.5 nm, D) i-layer, height: 350.5 nm, E) n-layer, height: 22.5 nm, F) AZO layer, height: 100 nm, G) silver layer, height: 200 nm, H) ARC (silicon nitride), height: 60 nm. Silver nano-disk is embedded in the silicon nitride ARC.

The critical parameters for the a-Si:H PSC design are the parameters related to the top metasurface. They are the diameter of the nano-disk '$d$,' the height of the nano-disk '$h$,' the period of the unit cell '$p$,' and the height of the anti-reflection coating (ARC) embedding the nano-disk '$h_{ARC}$.' These parameters are depicted in the PSC in Fig. 2. In this work, we used DE optimization algorithm to optimize these four parameters. The goal of the DE is to maximize the *OE* in the case of PSC. The bounded constraints for the metasurface parameters are described in Table I and the fitness function for DE to optimize is defined as follows:

$$\text{Maximize}: OE = \left(\frac{P_{i\text{-}a\text{-}Si:H(PSC)}}{P_{i\text{-}a\text{-}Si:H(\text{Re}f)}} - 1\right) \times 100 \qquad (9)$$

TABLE I. a-Si:H PSC parameters to be optimized and their constraints

| Parameters | Constraints |
|---|---|
| $d$ | 50 nm $\leq d \leq$ 300 nm |
| $h$ | 20 nm $\leq h \leq$ 80 nm |
| $p$ | 100 nm $\leq p \leq$ 800 nm |
| $h_{ARC}$ | 30 nm $\leq h_{ARC} \leq$ 120 nm |

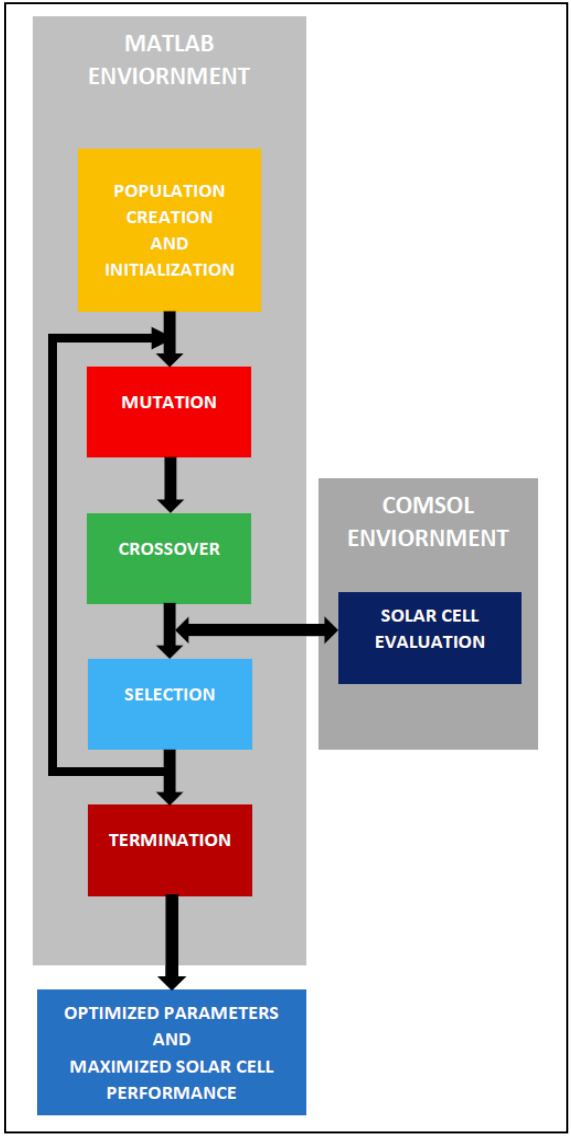

Fig. 3 DE optimization of a-Si:H PSC using COMSOL and MATLAB.

The control parameters for DE optimization algorithm are listed in Table II. Since the total number of parameters to be optimized is four ($D = 4$), the suggested $NP$ (population size) by Storn and Price [17] is between $5D$ and $10D$, where $D$ is the dimensionality of the optimization problem. In this work, we selected a reasonable population size of 32, which is $8D$. Additionally, the mutation scaling factor and the crossover probability was selected as 0.5 and 0.2 respectively, based on the work by Zhao et al. [18]. Also, the crossover strategy used in this work is *DE/rand-to-best/1/exp*. The maximum generations, $G_{max}$, to arrive at a converged solution or to terminate the iterations were selected as 500, as suggested by Zhao et al. [18].

TABLE II. DE control parameters

| DE control parameters | |
|---|---|
| Population size '$p$' | 32 |
| Scaling factor '$F$' | 0.5 |
| Crossover rate '$C_r$' | 0.2 |
| Crossover strategy | *DE/rand-to-best/1/exp* |
| Maximum generation '$G_{max}$' | 500 |

The DE optimization algorithm was implemented in COMSOL Multiphysics RF module v5.3b using LiveLink™ for MATLAB. The block diagram for the a-Si:H based PSC optimization using DE is depicted in Fig. 3. The DE is commenced on MATLAB, followed by the creation of population, initialization, and iterative steps of mutation, crossover, and selection. After the steps of mutation and crossover, which are executed in MATLAB environment, and the execution of trial vector parameters is done in COMSOL. After the model is simulated in the COMSOL environment, the absorbed power density is calculated, and the selection operation uses this value in DE, which is carried out in MATLAB. If the trial vector produces parameters which generate a design with higher absorbed power density, then the corresponding trial vector is replaced by the target vector, and the iteration continues until the termination criterion is satisfied.

The DE optimization algorithms were executed on 2 Intel Xeon E5 processors with 16 cores running at 2.50 GHz and 64 GB of RAM. In this DE optimization experiments, the execution time for evaluating fitness for one individual is approximately 2 min. The total computation times can be extrapolated from the population size ($NP$), maximum generation ($G_{max}$) and solution time per individual. Therefore, the expected time for DE to converge or DE to terminate is about 512 hours.

III. RESULTS AND DISCUSSION

In our previously reported work, we presented a design approach to maximize the absorption in i-a-Si:H layer of commercial a-Si:H solar cells and simultaneously reduce the undesirable effects of SWE by replacing the top contact layer (ITO) with 2D arrays of silver nano-disk patterned structures embedded into a silicon nitride ARC. One of the major challenges encountered while designing the PSC is device optimization, which increases the volume of the search space exponentially with an increase in parameters. This problem is well known as the "curse of dimensionality" in the optimization problem [19]. The DE optimization algorithm can significantly aid in reducing the manual efforts involved in optimized parameters search by automating the whole process with high efficiency and reducing the computational cost. The results of DE optimization algorithms on a-Si:H PSC are described in Table III. It can be observed that DE outperformed the previously reported a-Si:H PSC in performance and computational search time. The converged output of DE resulted in the optimum design of a-Si:H PSC with an absorbed power density of 321.2 W/m$^2$ and *OE* of 19.45 %, which is about 1 % improvement with respect to the reference compared to the previously reported design structure. Also, the DE optimized the solar cells in less than one-third of the time it took while optimizing manually.

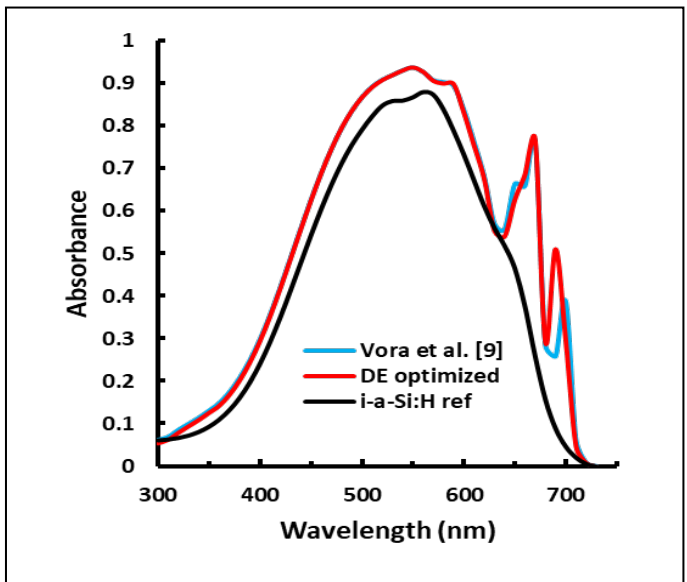

Fig. 4 Optical absorption in a-Si:H PSCs, (DE optimized and manually optimized [9]) and reference structures.

Table III. Comparing manual and DE optimization results

| Parameters | DE optimized | Manually optimized [9] |
|---|---|---|
| $d$ | 230 nm | 240 nm |
| $h$ | 51 nm | 50 nm |
| $p$ | 539 nm | 550 nm |
| $h_{ARC}$ | 60 nm | 60 nm |
| $P_{i\text{-}a\text{-}Si:H}$ | 321.2 W/m$^2$ | 318.7 W/m$^2$ |
| *OE* | 19.45 % | 18.51 % |
| Search time | 326 hours | >1500 hours |

Optical absorption in a-Si:H PSCs, both DE optimized and manually optimized in our previous work [9], and reference structures are depicted in Fig. 4. It can be observed that absorption spectra for both PSCs are nearly identical, except the one optimized using DE is relatively better in performance around 670 nm and 690 nm of incident sunlight. Also, the slight blue shift of resonance around 690 nm compared to the previously reported structure can be attributed to the shrinking of nano-disk diameter (Table III). Another interesting observation from Table III is the height of nano-disks and the height of ARC are identical for both PSCs. The height of ARC is best optimized at 60 nm irrespective of plasmonic nanostructures physical attributes, and therefore it came out to be identical while optimizing manually and with DE. Also, the height of the nano-disk was found to be identical since a thicker nano-structure can aid in exchanging the Ohmic losses in metallic nano-disk to the semiconductor layers thus improving the absorption characteristics. We also found that these two parameters converged much earlier in the initial phase of DE optimization and thus effectively reduced the dimensionality of the problem from 4 to 2, which significantly saved the computation time and escalated the convergence.

The convergence curves for DE optimization of a-Si:H PSC structures are presented in Fig. 5. It can be seen that DE algorithm converged at 422$^{nd}$ generation, much before the maximum number of generations are reached. We expected the DE algorithm to converge before 500 generations ($G_{max}$) which could take about 512 hours based on extrapolation as discussed earlier. However, the DE already converged at the 422$^{nd}$ generation which took approximately 326 hours of computational time. The DE was automatically terminated after 500$^{th}$ generation, which took about 437 hours of computation time. Comparing this search strategy to the manual search and optimization (Table III), it is evident that DE-based device optimization is not only fast, but also saves a lot of computation time and human resources.

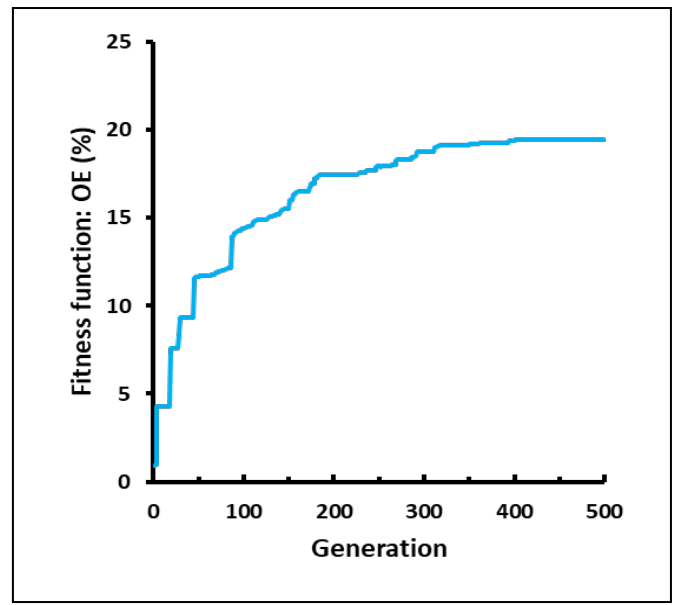

Fig. 5 Convergence curves for DE optimization of a-Si:H PSC structures

From a performance and design standpoint, the DE-based PSC optimization delivered a solution that met technical goals in maximizing the *OE* and demonstrated robustness and efficiency in parameters search. Fabrication limitations might require alterations to any final design or limit theoretically computed efficiencies. However, it is not complicated to introduce perturbations such as the introduction of surface morphology, fabrication tolerance, and complex geometries into the COMSOL model coupled with DE to generate a more accurate result. Secondly, from a computational standpoint, the successful implementation of a DE routine into complex electromagnetic design problem introduces significant savings for the user, by precluding the need for lengthy parametric sweeps. This is an important point since the DE automatically picks the optimal design that works for the desired frequency spectrum and geometrical boundaries without having to sweep the spectrum or a range of physical parameters. We did not perform a quantitative study on time savings or efficiency of DE compared to other black-box optimization algorithms, but this would be valuable future work for the benefit of metamaterials and plasmonics community. The DE optimization can also be used for scatterer geometries based solar cells that are difficult to predict for their performance and often limited to analytical models. In addition, an extension of this work will be to study and explore a more complex a-Si:H PSC which is a metasurface with hexagonal plasmonic arrays using nanoscale lithography [20].

## IV. Conclusion

This paper demonstrated a robust DE optimization algorithm implemented to assess the optical absorption and enhancement in the a-Si:H PSC and optimize the solution spaces for the design of top metasurface for performance enhancement. Computed *OE* for the a-Si:H PSC was found to be 19.45% above the reference structures, which is an improvement over the previously reported a-Si:H PSC design. The DE-based PSC device evaluation and optimization were found to be instrumental in demonstrating that a DE routine can generate and outperform manual search and maximize process for complex 3-D plasmonic structures in a time-efficient manner. DE-based PSC optimization is not only time efficient, but fully automated, robust and insightful. This method can be further extended to evaluate and improve the performance of complex 3-D plasmonic and optoelectronic device architectures for other applications that are limited by analytical models.

## Acknowledgment

The authors acknowledge National Science Foundation for supporting this work under the grant Award No. CBET-1235750. JMP acknowledge the support by Fulbright Finland for this work.